\documentclass[a4paper,twocolumn,10pt,accepted=2024-08-07]{quantumarticle}
\pdfoutput=1
\usepackage[utf8]{inputenc}
\usepackage[english]{babel}
\usepackage[T1]{fontenc}
\usepackage{amsmath}
\usepackage{hyperref}
\usepackage{tikz}
\usepackage{lipsum}
\PassOptionsToPackage{compress}{natbib}
\usepackage[numbers]{natbib}
\usepackage{listings}
\usepackage[resetlabels]{multibib}

\usepackage{hyperref}

\usepackage{dsfont}
\usepackage{amsfonts}
\usepackage{bbm}
\usepackage{bbold}
\usepackage{mathtools}
\usepackage{amsthm}
\usepackage{bm}
\usepackage{amssymb}
\usepackage[normalem]{ulem}
\usepackage{braket}
\usepackage{hyperref}
\usepackage{float}
\usepackage{tikz}
\usepackage{ifthen}
\usepackage{stmaryrd}
\usetikzlibrary{tikzmark}
\usepackage{upgreek}
\usepackage{braket}
\usepackage{physics}

\newcommand{\id}{\mathbbm{1}}

\begin{document}

\title{Quantum Chaos and Coherence: Random Parametric Quantum Channels}

\author{Apollonas S. Matsoukas-Roubeas}
\affiliation{Department of Physics and Materials Science, University of Luxembourg, L-1511 Luxembourg, G. D. Luxembourg}
\orcid{0000-0001-5517-0224}
\author{Toma\v{z} Prosen}
\affiliation{Faculty of Mathematics and Physics, University of Ljubljana, Jadranska ulica 19, 1000 Ljubljana, Slovenia}
\orcid{0000-0001-9979-6253}
\author{Adolfo del Campo}
\affiliation{Department of Physics and Materials Science, University of Luxembourg, L-1511 Luxembourg, G. D. Luxembourg}
\affiliation{Donostia International Physics Center, E-20018 San Sebasti\'an, Spain}
\orcid{0000-0003-2219-2851}

\begin{abstract}
The survival probability of an initial Coherent Gibbs State (CGS) is a natural extension of the Spectral Form Factor (SFF) to open quantum systems. 
To quantify the interplay between quantum chaos and decoherence away from the semi-classical limit, we investigate the relation of this generalized SFF with the corresponding $l_1$-norm of coherence.
As a working example, we introduce Parametric Quantum Channels (PQC), a discrete-time model of unitary evolution mixed with the effects of measurements or transient interactions with an environment.
The Energy Dephasing (ED) dynamics arises as a specific case in the Markovian limit. We demonstrate our results in a series of random matrix models.
\end{abstract}

\section{Introduction}
Realizable complex quantum systems are always subject to decoherence. 
Research over the last decades has led to the independent development of the theory of quantum chaos in isolation from environmental effects, \cite{gutzwiller1990,haake2018}, and that of open quantum systems  \cite{breuerbook,rivas2012}. 
Nevertheless, a common framework for their understanding without relying on semi-classical methods \cite{braun2001}, a theory of open quantum chaos, is yet to be developed.

Decoherence plays a crucial role in the emergence of classical-like chaotic behavior in quantum systems, such as phase-space islands and divergent trajectories governed by a positive Lyapunov exponent \cite{ZurekPaz94,Karkuszewski02,zurek2003,habib2005}.
Nonetheless, as the notion of chaos in quantum mechanics evolved, the spectral characterization of the system's Hamiltonian became the primary focus \cite{berry1977,bohigas1984,haake2018}.
The Bohigas-Giannoni-Schmit conjecture uncovered a relation between the distribution of energy levels in quantum analogs of classically chaotic systems and Random Matrix Theory (RMT) \cite{bohigas1984}, allowing the use of simple diagnostic tools, such as level-spacing distributions, for the heuristic classification of quantum chaotic systems, without reference to the semi-classical limit.
Soon, the application of Fourier methods shifted the identification of signatures of quantum chaos from the energy to the time domain \cite{leviandier1986,wilkie1991,alhassid1992,ma1995}. 
Notably, the analysis of the SFF revealed the importance of a ``correlation hole'', reflecting the effect of level repulsion.  
The SFF is closely related to the survival probability, the Loschmidt echo \cite{gorin2006}, quantum work statistics \cite{chenu2019},  frame potentials and t-designs \cite{collins2010,collins2015}, the partition function with complex temperature \cite{cotler2017,delcampo19,xu2019}, and the full-counting statistics of the Hamiltonian \cite{CaoXu22}. 
The recognition of the role of statistical correlations associated with quantum chaos has motivated additional proposals for diagnostics directly related to entanglement \cite{wang2004,prosen2007,chaudhury2009,neill2016,vidmar2017,torres_herrera2017,kumari2019,dowling2022} and quantum discord \cite{madhok2013}.
In this context, coherence in the energy eigenbasis is an essential ingredient of quantum chaos \cite{sieber2001}.

The unitarity of the dynamical maps generated by Hermitian Hamiltonians preserves the inner product between two arbitrarily close states. 
Consequently, quantum irregular motion cannot be characterized by a direct analog of the Lyapunov exponent or any other measure of sensitive dependence on initial conditions.
An intuitive alternative is the study of the sensitivity of the evolution to slight changes in the dynamical parameters, i.e., the Hamiltonian itself, quantified by the Loschmidt echo \cite{gorin2006}.
More recently, the fidelity susceptibility \cite{sierant2019,sels2021} has attracted considerable attention as a measure of eigenstate sensitivity to deformations.
Out-of-Time-Order Correlators (OTOC) \cite{larkin1969,hashimoto2017,maldacena2016} are promising candidates for the link between the phenomenology of eigenstate thermalization \cite{deutsch1991,srednicki1994,rigol2008,dalessio2016,bohrdt2017,murthy2019}, quantum information scrambling \cite{hayden2007,sekino2008,roberts2017,mi2021} and operator growth \cite{nahum2018,vonkeyserlingk2018}.
Lately, efforts have targeted the direct generalization of OTOC to study the fate of these effects in open systems \cite{reichental2018,Zanardi2021,schuster2022}.
Each measure probes complementary aspects of quantum chaotic behavior \cite{Bhattacharyya2022}.



In isolated systems, the Hermiticity of the Hamiltonian guarantees a set of real eigenvalues whose spacing distributions can exhibit level repulsion. 
In close analogy, an approach to open quantum chaos relies on characterizing the spectrum of Liouvillians and quantum channels.
Nevertheless, one cannot uniquely order the differences between their eigenvalues since they are complex numbers, in general. 
Therefore, the comparison of the corresponding distributions requires extra assumptions.
Such static signatures are pursued on different fronts, including the radial repulsion of complex eigenvalues \cite{akemann2019}, steady state eigenvalue distributions \cite{PZ2013,ribeiro2019,sa2020,sa2020a}, complex spacing ratios \cite{sa2020b}, and symmetry classification \cite{Ashida20,Hamazaki20,garciagarcia2022,sa2022symmetry,kawabata2022}. 
While some of the above studies preserve the simplicity and elegance of the Hermitian setting, they are less suited for quantifying the interplay between quantum chaos and decoherence, on which we focus. 

A complementary approach we pursue in this article relies on the generalization of the notion of SFF in an information-theoretic interpretation, i.e., as the Uhlmann fidelity \cite{uhlmann1976} between an initial state -- the thermofield double or a coherent Gibbs state -- and the time-evolving state, with no restrictions on the underlying dynamics \cite{delcampo17,xu2021,Cornelius22,matsoukas_roubeas2023}. 
This generalization of the SFF to open systems differs from alternative proposals focused on the spectral properties of the generator of the dynamics \cite{li2021,VikramGalitski22,ZhouZhouZhang23}. 
Within this approach, decoherence  generally suppresses the dynamical signatures of Hamiltonian quantum chaos. 
Namely, with the increase of dissipation, the depth and area of the correlation hole shrink, and the span of the ramp decreases \cite{delcampo17,xu2021,Cornelius22,LiaoGalitski22,matsoukas_roubeas2023}. 
This feature is also present in generalizations of the SFF based on the characteristic function of the local energy distribution in multipartite systems \cite{CaoXu22}. 
An interesting counterexample to this generic observation is that of balanced norm gain and loss in ED processes, where the duration of the linear ramp before the plateau is extended and the depth and area of the correlation hole can increase with finite dissipation \cite{Cornelius22,matsoukas_roubeas2023}.

In such context, relating coherence quantifiers to signatures of quantum chaos is an intricate task. The comparison of coherence quantifiers, such as coherence monotones, the $l_2$-norm of coherence \cite{streltsov2017}, and the coherence-generating power \cite{zanardi2017,styliaris2018,styliaris2019}, presents its own set of challenges.
Different coherence quantifiers often share qualitative similarities and can sometimes be bounded by one another. 
The authors of \cite{anand2021} have used quantifiers of coherence to diagnose quantum chaos in isolated dynamics; however, such quantifiers were not coherence monotones.
In this work, we seek to investigate how signatures of quantum chaos related to spectral statistics are altered when the evolution becomes nonunitary. 
In particular, we focus on the relation of the SFF to a coherence monotone, the $l_1$-norm of coherence.

We consider a quantitative bridge between the spectral correlations witnessing quantum chaos and decoherence by studying the survival probability, the purity, and the $l_1$-norm of coherence of an initial CGS.
At this level, the connection to the SFF becomes explicit.
This analysis reveals how the decay of the density matrix's off-diagonal elements in the energy eigenbasis directly suppresses the manifestation of level repulsion in the correlation hole.
Such a line of thought links quantum chaos in open systems to the resource theory of coherence \cite{streltsov2017,streltsov2018,chitambar2019,winter2016}. 
We note that the latter has recently motivated the use of coherence quantifiers as diagnostic tools for quantum chaos \cite{anand2021}.

To illustrate the generality of our approach, after discussing the case of ED dynamics, we introduce PQC, a family of mixed quantum operations whose consecutive application results in principle in non-Markovian quantum dynamics \cite{wolf2008a,rivas2014}.
Specifically, by splitting the dynamical map into two weighted channels, we introduce dissipation through the recurring effect of a set of Kraus-Choi operators on the isolated dynamics generated by a Hamiltonian.
This model is of interest as a paradigm of mixed Completely Positive and Trace-Preserving (CPTP) operations involving a fixed fundamental period of interaction with a reservoir. So as to quantify the interplay between quantum chaos and decoherence, we analyze the spectral signatures and the correlation hole in the SFF for varying values of the control parameters, sampling the constituent matrices of the channel from random matrix ensembles. 

The paper is structured as follows. 
Sec. \ref{SecSFFopen}  introduces the fidelity-based definition of the SFF as a relevant extension of the notion to open systems.
Sec. \ref{SecDTFD}  reviews the definitions of the purity and the most relevant coherence monotones of the CGS, relating them to the SFF.
Sec. \ref{SecED} addresses the case of decoherence in the energy eigenbasis when the Hamiltonian and the dissipator commute.
Sec. \ref{SecRPQC}  discusses the properties of random PQC.

\section {Spectral Form Factor for Open Quantum Systems} \label{SecSFFopen}

The survival probability of a quantum state evolving under unitary dynamics carries information about the structure of the energy levels and their correlations.
It equals the fidelity between a pure initial and the time-evolving state, while its Fourier transform is a weighted sum of Dirac $\delta$-functions positioned at the Hamiltonian eigenvalues. 

The role of a so-called correlation hole, in the survival probability and its connection to level statistics have appeared early on in the literature \cite{leviandier1986,guhr1990,wilkie1991,alhassid1992,ma1995}.
The correlation hole is a dip below the saturation value, followed by a linear ramp that concludes at the time when the smallest dynamical frequencies have been expressed.
The depth of the correlation hole is a measure of eigenvalue repulsion, typical of Hamiltonian quantum chaos, which could be carried over into dissipative systems.

Let us consider a Hilbert space of finite dimension $d \in \mathbb{N}$,  a Hermitian Hamiltonian $H = \sum_{n=1}^d E_n \ketbra{n}{n}$,  the corresponding partition function $Z(\beta)=\tr[\exp(-\beta H)]$  and the Boltzmann weights $p_n = \exp(-\beta E_n) / Z(\beta)$.
The CGS at inverse temperature $\beta = (k_B T)^{-1}$ is defined as 
\begin{align}
\ket{\Psi_\beta} = \sum_{n=1}^d  \sqrt{p_n}  \ket{n} .
\end{align}
For isolated dynamics, the SFF coincides with the fidelity between an initial CGS and its time evolution \cite{delcampo17,chenu2019,Martinez-Azcona22}
\begin{align} \label{isoSFF}
{\rm SFF}_\beta(t) =  \abs{ \bra{\Psi_\beta}  e^{- i \frac{t}{\hbar}  H} \ket{\Psi_\beta} }^2 = \abs{\frac{Z(\beta + i t)}{Z(\beta)}}^2.
\end{align}
The generalization of the above definition for open quantum systems has been put forward in \cite{xu2021} and further studied in \cite{Cornelius22,matsoukas_roubeas2023,ZhouZhouZhang23}.
These results indicate that decoherence generally suppresses the depth and area of the correlation hole.
Nevertheless, the time duration of the ramp can be extended when the quantum jumps are neglected \cite{Cornelius22,matsoukas_roubeas2023}. 
Here, we focus on the dynamics generated by a CPTP operation. 
In general, given an arbitrary CPTP quantum channel $\Upphi_t$, depending on a discrete or continuous time parameter $t$, the fidelity-based SFF reads \cite{xu2021},
\begin{align} \label{sffopen}
{\rm SFF}_\beta(t) = \bra{\Psi_\beta}   \rho_{\beta}(t) \ket{\Psi_\beta},
\end{align}
with $\rho_{\beta}(t)= \Upphi_t [ \ketbra{\Psi_\beta} ]$.
We note that this is equivalent to the standard notion of the SFF for unitary maps while applicable to any other quantum evolution involving a Hamiltonian, such as circuit models, dissipative Floquet systems, etc.
Nonetheless, at infinite temperature, $\beta=0$, all the Boltzmann factors are equal $p_n=1/d$ hence a reference Hamiltonian is not necessary for the above definition, which is reduced to
\begin{align} \label{sffopenbeta0}
{\rm SFF}(t) = \frac{1}{d} \left( 1 +  2 \sum_{\substack{n,m=1 \\ m < n}}^d \Re[\rho_{0, nm} (t)] \right) ,
\end{align}
writing for simplicity ${\rm SFF}(t) \equiv {\rm SFF}_0(t)$, since our calculations in this article will mainly refer to the infinite temperature CGS.
In Sec. \ref{SecED}, the quantum channel will be generated by a Linbladian having as only dissipator the system Hamiltonian itself, while in Sec. \ref{SecRPQC}, we construct a heuristic discrete-time CPTP map in which unitary evolution is mixed with arbitrary physical processes. 

In what follows, we will often adopt a specific matrix representation of all superoperators for the intuitive simplification of their spectral properties, often referred to as the Liouville space formalism \cite{gyamfi2020}.  
In particular, we consider the horizontal vectorization of any density matrix
\begin{align}
 \rho =  \sum_{n,m=1}^d \rho_{nm} \ketbra{n}{m}  
\rightarrow | \rho ) := \sum_{n,m=1}^d \rho_{nm}   \ket{n} \otimes \ket{m}^*   .
\end{align}
Once the density matrix is decomposed in such a representation, the choice of basis is fixed, and any further transformation of a superoperator has to be treated consistently.  
The horizontal decomposition allows the expression of the Hilbert-Schmidt inner product between two operators as the usual projection product between vectors $\Tr[A^\dagger B] = (A|B)$.
In addition, it allows the expression of linear superoperators acting on a state as a Kronecker product between the operator acting from the left and the transpose of the operator acting from the right, $A \rho B \rightarrow A \otimes B^{\intercal} | \rho )$.  
We use capital letters (e.g., $\Upphi,\Uplambda,\dots$) for the superoperators acting on a density matrix in the standard density operator formalism, and the corresponding blackboard bold (e.g.,  $\mathbb{\Phi},\mathbb{\Lambda},\dots$) for their matrix representation in the Liouville space formalism.
The horizontal vectorization of the CGS is
\begin{align}
| \rho_\beta ) = \sum_{n,m=1}^d \frac{e^{-\frac{\beta}{2}   ( E_n + E_m )}}{Z(\beta)}   \ket{n} \otimes \ket{m}^*   .
\end{align}
The proposed SFF for open systems \eqref{sffopen}, given the matrix representation of the quantum channel $\Upphi_t \rightarrow \mathbb{\Phi}_t = \{ \mathbb{\Phi}_{t,\ell k} \}_{\ell,k=1}^{d^2}$, becomes ${\rm SFF}_\beta(t) = ( \rho_\beta | \mathbb{\Phi}_t | \rho_\beta)$.
At infinite temperature, it is equal to the sum of all elements
 \begin{align} 
{\rm SFF}(t)= \frac{1}{d^2} \sum_{\ell, k =1}^{d^2} \mathbb{\Phi}_{t,\ell k}  .
\end{align}

\section{Coherence Monotones and Quantum Chaos}\label{SecDTFD}

Dephasing is a form of decoherence that suppresses the density matrix's off-diagonal entries in a given basis.
When examining thermalization and quantum ergodicity, the diagonal ensemble motivates the Hamiltonian eigenbasis as the most relevant for the study of decoherence.
In the energy eigenbasis, coherence quantifiers can capture how localized or uniformly spread a quantum state is, i.e., they serve as delocalization measures \cite{anand2021}. 

The CGS and the closely related thermofield double state are natural probe states for quantum chaos.
They are coherent quantum states with support in the whole energy spectrum and a tunable filter (the square root of the Boltzmann weights), preferentially sampling low-energy states at finite inverse temperature $\beta>0$. 
Furthermore, in the resource theory of quantum coherence, the infinite temperature CGS, $\beta=0$, is a maximally coherent state, saturating the value of any well-defined coherence monotone. 
In this context, any quantum state can be prepared from the infinite temperature CGS, using only incoherent operations \cite{baumgratz2014,levi2014,streltsov2017}.

Motivated by the central role of the CGS in the study of quantum chaos and resource theory of coherence, we propose the known coherence monotones of its time evolution as relevant diagnostic tools for the interplay between the two.
We introduce a relation between the SFF and the $l_1$-norm of coherence, suggesting the study of other relevant signatures of quantum chaos or coherence monotones for future investigations.

The $l_1$-norm of coherence is a coherence monotone \cite{baumgratz2014}, defined as
\begin{equation}
 \label{cl1}
C_{l_1} (t) =  2 \sum_{\substack{n,m=1 \\ n < m}}^d \abs{ \rho_{n m} (t) }    .
\end{equation}
The purity of a density matrix, $P = \Tr[ \rho^2] $, is in any basis equal to the sum of the squares of all elements
\begin{align}
 P (t) = \sum_{\ell=1}^d \abs{ \rho_{n n}(t) }^2 +
 2 \sum_{\substack{n,m=1 \\ n < m}}^d \abs{ \rho_{n m}(t) }^2 .
\end{align}
By observing that $ 0 \leq \abs{ \rho_{n m}(t) } \leq 1$, for all $n,m = \{ 1,2,\dots,d \}$, one can directly relate the two through the inequality
\begin{equation} \label{purcl1}
P (t) \leq  \sum_{n=1}^d \abs{\rho_{n n}(t)}^2 + C_{l_1} ( t)  .
\end{equation}
We note that the first term on the right-hand side of the last inequality can be associated with the inverse participation ratio of the initial state in the computational basis, which can serve as a measure of delocalization. Further, it equals the purity the decohered density matrix $\sum_n|n\rangle\langle n|\rho(t)|n\rangle\langle n|$.
In what follows, we will study the properties of purity and the $l_1$-norm of coherence and their relation to fidelity when the initial state is the CGS.   

At infinite temperature, $\beta=0$, the SFF is bound by the $l_1$-norm of coherence of the CGS through
\begin{equation} \label{sffleqcl1}
\frac{1 - C_{l_1}(t)}{d} \leq  {\rm SFF}(t)  \leq \frac{1 + C_{l_1}(t) }{d}   .
\end{equation}

For a detailed proof, please see appendix \ref{apx5}.
The above bound helps to understand the extent and timescales at which quantum fluctuations are suppressed by decoherence, distinguishing this effect from their reduction due to ensemble averaging.
We note that the authors of Ref. \cite{anand2021} relate the SFF to the coherence-generating power \cite{zanardi2017,styliaris2018,styliaris2019} averaged over the Gaussian unitary ensemble through an analytical upper bound.
Nevertheless, the coherence-generating power is a property that describes a quantum operation and its ability to generate coherence from incoherent states, rather than describing the states themselves.
We do not address the possibility of deriving inequalities similar to \eqref{sffleqcl1},  with different coherence monotones and measures, such as the relative entropy of coherence \cite{streltsov2017}. 

\section{Energy Dephasing} \label{SecED}
The time-continuous description of dephasing in the energy eigenbasis is a natural model of decoherence known as ED. It arises in scenarios involving timing of the quantum unitary dynamics with noisy clocks \cite{Egusquiza99}, the theory of fluctuating Hamiltonians \cite{Adler03}, time coarse-graining, continuous measurement of energy, and stochastic modifications of quantum mechanics \cite{Gisin84,Milburn91,Percival94,Bassi13}, to name some relevant examples. Due to its simplicity, it is amenable to analytical description and has been used in open quantum chaos and black hole physics \cite{xu2019,delcampo19,xu2021,Cornelius22,matsoukas_roubeas2023}.
The ED dynamics is governed by the equation
\begin{equation}
\label{dyned}
\partial_t \rho (t) = -\frac{i}{\hbar} [H,\rho (t)]-\gamma[ H,[H,\rho (t)]],
\end{equation}
with dephasing strength $\gamma$. 
The general solution reads
\begin{equation}
\rho (t)=\sum_{n,m=1}^d \rho_{nm}(0)e^{-\frac{i}{\hbar}  t(E_n-E_m)-\gamma t(E_n-E_m)^2} \ketbra{n}{m}.
\end{equation}

For an initial CGS, $\rho_{nm}(0)=\sqrt{p_np_m}$ and the SFF  becomes
\begin{eqnarray}
\label{sffed}
&&{\rm SFF}_\beta(t)  
= F_p \\
&&+ 2 \sum_{m < n} p_n p_m    e^{-\gamma t (E_n-E_m)^2}\cos \bigg(\frac{E_n-E_m}{\hbar}    t \bigg),
\nonumber
\end{eqnarray}
where $F_p=Z(2 \beta)/Z^2(\beta)$ is the value of the plateau reached after the Heisenberg time.
Accordingly, the $C_{l_1}$ of the CGS under dephasing is
\begin{equation}
\label{cl1ed}
C_{l_1,\beta} ( t ) =   2 \! \sum_{m < n}  \sqrt{ p_n p_m}   e^{- \gamma t(E_n-E_m)^2},
\end{equation}
while the corresponding purity  reads
\begin{equation}
\label{purityed}
P_\beta (t) = F_p  + 2 \! \sum_{m < n}  p_n  p_m    e^{- 2 \gamma t(E_n-E_m)^2} .
\end{equation}

We note that the SFF and purity of the CGS saturate at the same plateau, $F_p$, while the $l_1$-norm of coherence goes to zero.
The reason is the following.
Incoherent states are always diagonal in the reference basis, $\rho=\sum_{n=1}^d r_n \ketbra{n}$, with probabilities $r_n$ \cite{streltsov2017}.
Here, the reference basis is the energy eigenbasis, and the dissipative part of Eq.~\eqref{dyned} commutes with the Hamiltonian; consequently, coherence can never be generated.
We note that ED is a maximally incoherent and translationally-invariant operation in the sense of \cite{streltsov2017}, mapping all incoherent states into incoherent states while asymptotically consuming all coherence of an initial coherent state such as the CGS.

\begin{figure}
\hspace*{0 cm}
\includegraphics[width=0.95 \linewidth]{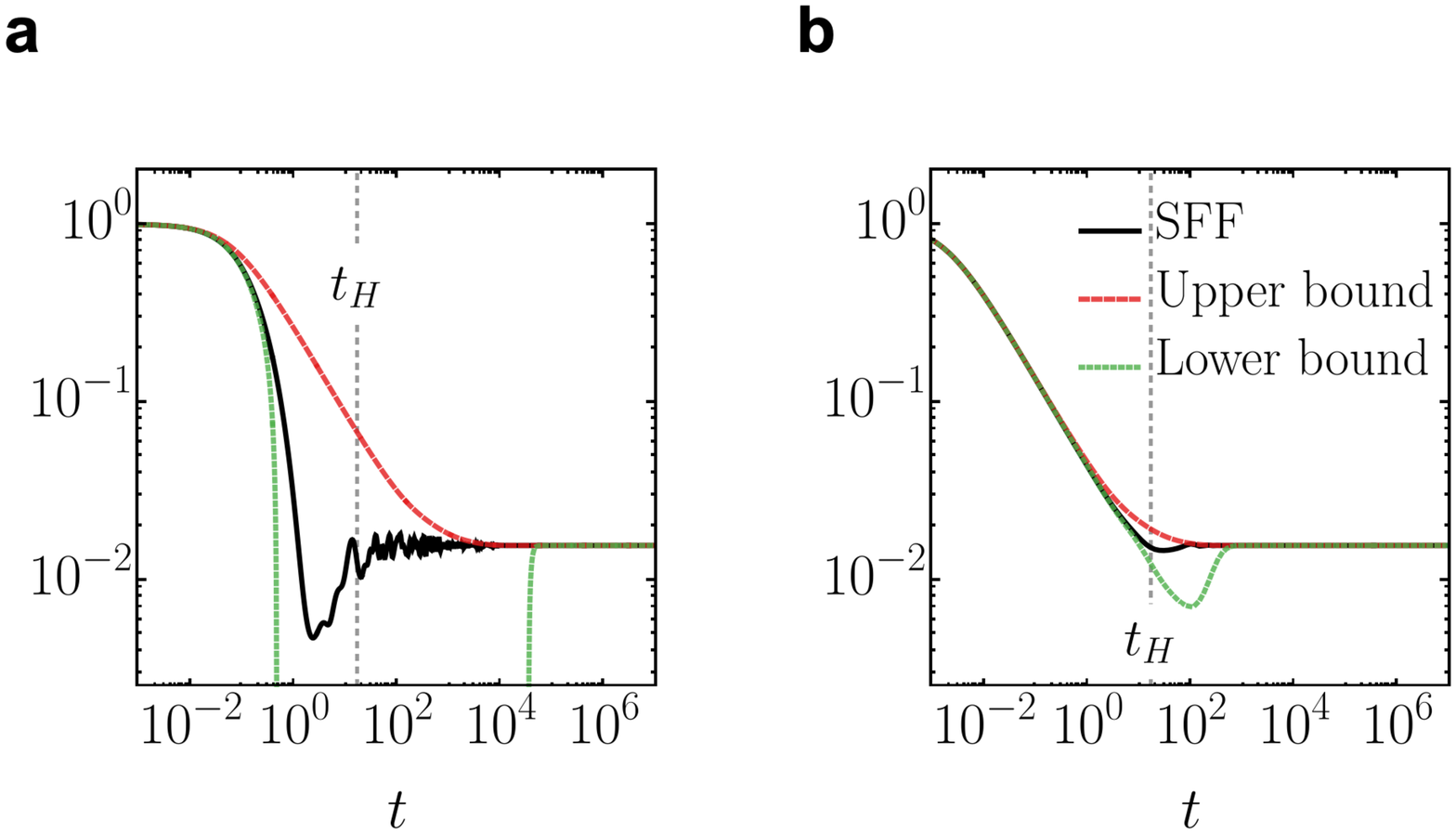}
\caption{
The correlation hole is a feature in the SFF of quantum chaotic systems, characterized by a dip below the saturation value, followed by a linear ramp, and concluding with a plateau.
The SFF is bounded by the $l_1$-norm of coherence of the CGS, from above by the general Eq.~\eqref{sffleqcl1} and from below by Eq.~\eqref{sffgeqcl1} when ED \eqref{sffed} is considered. 
Here, we show the corresponding numerical ED calculation for a single $\mathrm{GOE}(64)$, $\sigma=1$ Hamiltonian,
increasing the dephasing strength from $\gamma=0.1$ in panel {\sf{\textbf{a}}} to $\gamma=4$ in panel {\sf{\textbf{b}}}. The bounds become tight under strong ED when the correlation hole is suppressed but are weak otherwise. 
}
\label{PQCfig1}
\end{figure}

In isolation ($\gamma=0$), the Heisenberg time is defined as $ t_\text{H} = 2 \pi \hbar / \Delta$, with $\Delta$ being the mean level spacing of the Hamiltonian $H$.
It determines the onset of the plateau of the SFF and sets a limit to the duration of any control protocol of quantum information. 
After this timescale, all incommensurable frequencies governing the dynamical evolution of a complex quantum system have been expressed on average, canceling each other effectively, regardless of their contribution to the initial state.
In Fig. \ref{PQCfig1}, we see that the onset and value of the plateau remain unchanged under ED dynamics.
See also Fig. \ref{PQCfigapx3a} in appendix \ref{apx3} for a comparison of different $\gamma$.
By contrast, the minimum of the correlation hole, associated with the Thouless time  \cite{schiulaz2019}, is delayed.
 
For the purity and the $l_1$-norm of coherence of a CGS, we see from Eq.~\eqref{purcl1} that
\begin{equation}
P_\beta (t) \leq   F_p   + C_{l_1,\beta} ( t)   .
\end{equation}
At infinite temperature, $\beta = 0$, one can make a Taylor expansion of the cosine in Eq.~\eqref{sffed} and relate the SFF and the derivatives of $C_{l_1}$ with respect to $\gamma$ of the CGS through the equation
\begin{equation}
  {\rm SFF}(t) =  \frac{1}{d} \left( 1 + \sum_{j=0}^\infty  \frac{t^j}{ \hbar^{2j}   (2j)!}    \partial_\gamma^j    C_{l_1}(t) \right)  .
\end{equation}
The two first terms give the inequality
\begin{equation} \label{sffgeqcl1}
 {\rm SFF}(t) \geq \frac{1}{d} \left( 1 + C_{l_1}(t) + \frac{\partial_\gamma    C_{l_1}(t)}{2} t  \right) .
 \end{equation}
The above, together with Eq.~\eqref{sffleqcl1}, bound the fluctuations of the SFF around the plateau at large timescales.
Said differently, the volume of the "quantum noise" \cite{barbon2014} in the SFF is bound by the derivative of the $l_1$-norm of coherence of the CGS.

In Fig. \ref{PQCfig1}, we sample the Hamiltonian matrix from $\mathrm{GOE}(d)$, whose eigenvalue probability density function is the semicircle law 
\begin{align}
 \label{semicircle}
 \mu (E) = \frac{\sqrt{2 d \sigma^2 -E^2}}{\pi d \sigma^2},
\end{align}
denoting with $\sigma$ the standard deviation of the Gaussian distribution.
Before unfolding the spectrum, the mean level spacing for a Hamiltonian sampled from $\mathrm{GOE}(d)$ is $\Delta  = \frac{\sigma \sqrt{8 d}}{d-1}$.
As a result,  for ED dynamics, the Heisenberg time becomes
 \begin{align} \label{RMTheisenberg}
t_\text{H} = \frac{\pi \hbar (d-1)}{  \sigma \sqrt{2d} }.
\end{align}
The increase of the dephasing strength $\gamma$  gives rise to the gradual loss of the correlation hole \cite{xu2021}. The lower and upper bounds to the SFF derived in terms of the coherence monotone $C_{l_1}$ are weak in the presence of quantum coherences in the energy eigenbasis when the dynamical signatures of quantum chaos are manifest and become tighter as the dephasing strength is increased.

\section{Parametric Quantum Channels} \label{SecRPQC}

In the previous sections, we examined the quantitative relation between the SFF and the $l_1$-norm of coherence of an initial CGS. 
As a first example, we performed numerical calculations for the maximally incoherent operation of ED.
In this section, we define a non-Markovian model describing a binary mixture of quantum channels; one unitary, the other generic.

Let us consider the convex subspace of a Hilbert-Schmidt space $\mathcal{D(H)}$, containing all unit-trace, Hermitian and positive semi-definite density matrices $\rho \in \mathcal{D(H)}$, acting on a Hilbert space, $\mathcal{H} \subseteq \mathbb{C}^d$ of finite dimension $d \in \mathbb{N}$. 
Extending the $1$-parameter model \cite{sa2020}, we introduce the operator-sum representation of the PQC on $\mathcal{D}(\mathcal{H})$ through
 \begin{align}  
  \label{model}
\thinmuskip=0mu
\Uplambda_{\uptau,\epsilon} [\rho] = (1-\epsilon)   e^{-i \frac{\uptau}{\hbar}   H}   \rho   e^{ i \frac{\uptau}{\hbar}   H} + \epsilon   \sum_{r=1}^{K} N_r^{\phantom \dagger} \rho N_r^\dagger ,
 \end{align} 
with the parameters $\epsilon \in [0,1]$, $\uptau \in \mathbb{R}$, and $K \in \{1,2, \dots, d^2-2\}$.  
This $2$-parameter family of CPTP maps can be thought of as a prototype of unitary temporal evolution, mixed with the effects of measurements \cite{sierant2022} or transient interactions with an environment \cite{krausbook}.
We define the dissipation strength $\epsilon$ to be a small parameter that weights the non-unitary part of the channel, comprising arbitrary environmental degrees of freedom by the action of the Kraus-Choi operators $N_r \in \mathcal{HS}$
 \begin{align}  
  \label{tracepres}
\sum_{r=1}^{K}  N_r^\dagger    N_r^{\phantom \dagger} = \id   .
 \end{align} 
We explicitly include a dissipation period $\uptau$, which resolves the unitary propagator by rescaling the Hamiltonian $H$, bounding the eigenvalue distribution of the corresponding superoperator spectrum within a specific angle on the complex plane. 
As we will shortly see, whenever $\uptau \ll \hbar/\epsilon $, its inverse $1/\uptau$ can effectively approximate the frequency with which an effect interrupts the unitary evolution with a dissipative kick of strength $\epsilon$.
Thus, a relative timescale and a corresponding energy scale $\hbar / \uptau$ are set, indicating how fast the internal degrees of freedom change compared to the accessible environmental processes.

\subsection{Discrete Time Evolution}
The vectorized representation of a PQC of  Eq.~\eqref{model} becomes
\begin{align}
  \label{supermodel}
\thinmuskip=0mu
\mathbb{\Lambda}_{\uptau,\epsilon}  = (1-\epsilon)   e^{i \frac{\uptau}{\hbar}   (\id \otimes H^\intercal - H \otimes \id)} + \epsilon   \sum_{r=1}^{K} N_r^{\phantom *} \otimes N_r^*   .
\end{align}

Accordingly, $j \in \mathbb{N}$ consecutive applications of a PQC can be represented as
 \begin{align}
  \label{supermodelpower}
\thinmuskip=0mu
\mathbb{\Lambda}_{\uptau,\epsilon}^j  = \bigg( (1-\epsilon)   e^{i \frac{\uptau}{\hbar}   (\id \otimes H^\intercal - H \otimes \id)} + \epsilon   \sum_{r=1}^{K} N_r^{\phantom *} \otimes N_r^* \bigg)^j .
\end{align}
The generated discrete-time dynamics can provide a heuristic approach to a class of physical phenomena by the introduction of a time parameter $t = j   \uptau$,  generally irreproducible by a continuous master equation, $\dot{\rho} (t) = \mathcal{L}    \rho(t)$, with $\mathcal{L}$ being in the Lindblad form \cite{lindblad1976,gorini1976}.
In particular, if the channel $\mathbb{\Lambda}_{\uptau,\epsilon}$ is non-divisible, then the corresponding quantum evolution is non-Markovian \cite{wolf2008a,wolf2008b}. Nevertheless, whenever the channel $\mathbb{\Lambda}_{\uptau,\epsilon}$ satisfies necessary and sufficient conditions for the quantum process it describes to be Markovian \cite{pollock2018,milz2019}, the evolution of Eq.~\eqref{supermodelpower} can, in principle, be represented by a master equation of Lindblad form \cite{breuerbook,rivas2012}.

Indeed, in the limit of very small dissipation strength and large dissipation period, we can define $\epsilon = 2\gamma \uptau$ and the difference
 \begin{align}  
  \label{markov1}
  \rho(t+\uptau) -  \rho(t) &= \Uplambda_{\uptau,\epsilon} [ \rho(t) ] -  \rho(t) 
  \\ \nonumber
  \simeq  (1-\epsilon) \Big( \rho(t) &- i   \frac{\uptau}{\hbar}   [ H ,  \rho(t) ] \Big)
  + \epsilon   \sum_{r=1}^{K} N_r^{\phantom \dagger} \rho (t) N_r^\dagger 
  - \rho(t)   .
 \end{align} 
From the trace preservation Eq.~\eqref{tracepres}, one obtains the identity
 \begin{align}  
  \label{markov2}
 \rho(t) = \frac{1}{2} \sum_{r=1}^{K}  \big\{ N_r^\dagger    N_r^{\phantom \dagger} ,    \rho(t) \big\} .
 \end{align} 
Finally, the division of both sides of Eq.~\eqref{markov1} by the dissipation period, followed by the limit $\uptau \rightarrow 0$, yields the time-continuous master equation
 \begin{align}  
  \label{lindblad}
  \dot{\rho} (t) =    &- \frac{i}{\hbar}   [ H ,  \rho(t) ]  \\ \nonumber
  &+2 \gamma    \sum_{r=1}^{K} \left( N_r^{\phantom \dagger}    \rho (t)    N_r^\dagger 
  -  \frac{1}{2} \big\{ N_r^\dagger    N_r^{\phantom \dagger} ,    \rho(t) \big\} \right)    .
 \end{align} 
In this limit, the Kraus-Choi operators become the Lindblad operators.
As a remark, weak but frequent interactions with the environment can result in large $\gamma$. 
Moreover, the ED model discussed in Sec. \ref{SecED} can be derived from the PQC, choosing the system Hamiltonian as the single Kraus-Choi operator $N_1=H$. 
The master equation (\ref{dyned}) results from PQC in the continuum limit $\uptau\rightarrow0$ keeping $\gamma=\epsilon/(2 \uptau)$ constant.

In the model of  Eq.~\eqref{model}, the fidelity between the initial pure CGS and the state after $j$ applications of the PQC is
\begin{align}
{\rm SFF}_\beta (j\tau) = \bra{\Psi_\beta} \underbrace{\Uplambda \circ \dots \circ\Uplambda [  }_{\text{$j$-times}} \rho_\beta]\ket{\Psi_\beta}   ,   
\end{align}
or equivalently, in the vectorized notation 
 \begin{align}
{\rm SFF}_\beta (j\tau) = ( \rho_\beta | \mathbb{\Lambda}_{\uptau,\epsilon}^j | \rho_\beta)   .
\end{align}

\begin{figure}
\hspace*{0.4 cm}
\includegraphics[width=0.8\linewidth]{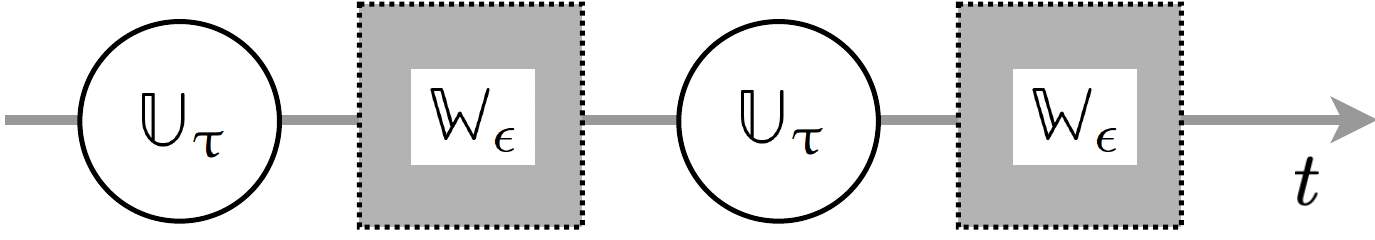}
\caption{Schematic representation of unitary evolution periodically interrupted by a 1-parameter channel.
}
\label{PQCfig2}
\end{figure}

We note that the discrete evolution of  Eq.~\eqref{supermodelpower} has a unique steady state whose eigenvalue distribution determines the asymptotic value of the purity at long times $P(t \rightarrow \infty)$. 
The PQC evolution can effectively result from a unitary quantum evolution that is periodically interrupted by the action of a dissipative quantum channel.   
Specifically, let us consider the sequential application of a unitary 
\begin{align}
  \label{unitary}
\mathbb{U}_\uptau  = e^{i \frac{\uptau}{\hbar}   (\id \otimes H^\intercal - H \otimes \id)}   ,
\end{align}
and an $1$-parameter channel
\begin{align} \label{wchannel}
\mathbb{W}_\epsilon = (1-\epsilon)   \id_{d^2} + \epsilon   \sum_{r=1}^{K} N_r^{\phantom *} \otimes N_r^*   ,
\end{align}
acting alternately $j$ times on an initial density matrix $| \rho_0 )$
\begin{align}
| \rho_ j )  = ( \mathbb{W}_\epsilon \mathbb{U}_\uptau )^j   | \rho_ 0 )   ,
\end{align}
as shown in Fig. \ref{PQCfig2}.
The channel of Eq. \eqref{wchannel} can be interpreted as a weak instantaneous perturbation from the environment for small $\epsilon$.
The time evolution at every step is described by
\begin{align} \label{chaninter}
\mathbb{W}_\epsilon \mathbb{U}_\uptau &= (1-\epsilon)   e^{i \frac{\uptau}{\hbar}   (\id \otimes H^\intercal - H \otimes \id)} \\ \nonumber
&+ \epsilon   \left( \sum_{r=1}^{K} N_r^{\phantom *} \otimes N_r^* \right)   e^{i \frac{\uptau}{\hbar}   (\id \otimes H^\intercal - H \otimes \id)}   .
\end{align}
The channels $\mathbb{\Lambda}_{\uptau,\epsilon}$ of  Eq.~\eqref{supermodel} and $\mathbb{W}_\epsilon \mathbb{U}_\uptau$ differ in the unitary $\mathbb{U}_\uptau$ multiplying the dissipative part. When all constituent matrices are sampled randomly, this plays a minor role in the spectral properties of the overall channel.
More generally, Eq. \eqref{supermodel} is a good approximation of Eq. \eqref{chaninter} if $\epsilon \uptau \ll \hbar$, as seen by truncating the Taylor expansion of the exponential multiplying its dissipative part.

\subsection{Random Quantum Channels}

The spectral signatures of complex Hamiltonian systems, carrying all information needed for the description of the evolution of an isolated system, have served for almost four decades as an archetype of quantum chaos \cite{berry1977,bohigas1984,haake2018,gutzwiller1990}.  
Consequently, the classification of self-adjoint and unitary ensembles has been a priority of mathematical physicists in the previous century.
In parallel, a concrete theory of open quantum systems has been developed \cite{lindblad1976,gorini1976,krausbook,breuerbook,rivas2012}, shifting the attention to the overall dynamical maps governing the temporal evolution of a subsystem.  
In the Lindbladian decomposition of the generators of such dynamical maps, the Hamiltonian operator becomes an element of a set of not-necessarily-Hermitian operators determining the dynamics.
From an information-theoretic perspective, a dynamical map is a completely positive superoperator mapping quantum states to quantum states.
The PQC introduced in  Eq.~\eqref{model} allows the explicit inclusion of a quantum chaotic Hamiltonian in the later picture while incorporating the decohering action of arbitrary environmental effects.

We next perform a series of random matrix calculations to unveil and quantify the behavior of various signatures of Hamiltonian quantum chaos in the presence of decoherence.
The first part of the PQC of Eq.~\eqref{model} encodes the standard paradigm of Hamiltonian quantum chaos \cite{haake2018,metha2004}, while the second part is taken as a composition of random Kraus-Choi operators, the weight of which represents the strength of the interaction with the environment.  
Specifically, we sample the Hamiltonian matrix $H$ from the Gaussian orthogonal ensemble, $\mathrm{GOE}(d)$ and construct the rest $K$ Kraus-Choi operators $N_r$ as block-truncations of an enlarged random $\mathrm{CUE}(K   d)$ element as follows \cite{zyczkowski2000,bruzda2010,bruzda2009,kukulski2021,sa2020}. 
Let $\mathbb{V} \in \mathbb{C}^{K   d \times K   d}$ be a unitary, implying that $\sum_{\ell=1}^{K   d} \mathbb{V}_{\ell i}^*   \mathbb{V}_{\ell j}^{\phantom *} = \delta_{ij} $. 
One can introduce the submatrices $N_r$ as the $K$ consecutive $d \times d$ blocks of the $(K   d) \times d$ matrix created by any $d$ columns of $\mathbb{V}$.
Suppose we interchange two columns of a unitary. In that case, we obtain another unitary, so we consider the $n$ to $n+d$, $1 \leq n \leq d(K-1)$ consecutive columns without any loss of generality, $N_{r,\nu \mu} \equiv \mathbb{V}_{(r-1)d+\nu,n+\mu }$, with $\nu, \mu={1,2,\dots,d}$.  
Then, $N_r$ are Kraus-Choi matrices respecting the trace preservation property \eqref{tracepres}. 

\subsection{Spectral Phases}

There is a growing interest in the classification of the spectral properties of random non-Hermitian Hamiltonian operators \cite{feinberg1997,wang2020a,mochizuki2020,marinello2016}, random Liouvillian operators \cite{sa2020a,wang2020}, random Lindblandian operators \cite{denisov2019,can2019,can2019a,lange2021,kawabata2022} and random Kraus maps \cite{sa2020,kukulski2021}.
We dedicate this subsection to the characterization of the spectral phase diagram of random PQC, providing analytic estimations of the spectral loci as functions of the dissipation parameters.

\begin{figure}
\hspace*{0.2 cm}
\includegraphics[width=1\linewidth]{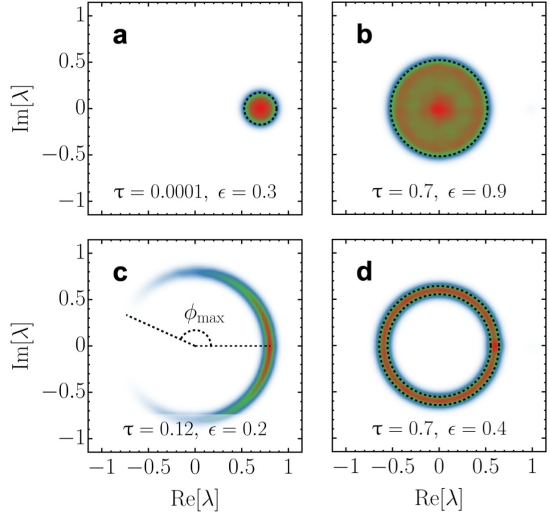}
\caption{Spectral phases of random PQC.
Density plot on the complex plane (blue to red) of $16384$ eigenvalues of the superoperator defined in  Eq.~\eqref{supermodel}, together with the theoretical boundaries (black dashed lines). 
Each panel was generated by $4$ independent realizations, with the Hamiltonian $H$ sampled from $\mathrm{GOE}(64)$, $\sigma =1$, setting $\hbar = 1$ and a set of $K=3$ random Kraus-Choi matrices $N_r$, drawn as truncations of an enlarged random $\mathrm{CUE}(192)$ element. 
{\sf{\textbf{a.}}} Shifted-disk phase: the boundary of the spectral locus is given by  Eq.~\eqref{sdboarder}.
{\sf{\textbf{b.}}} Disk phase: the boundary of the spectral locus is given by  Eq.~\eqref{dboarder}.
{\sf{\textbf{c.}}} Crescent phase (transitive area): the spectral locus is confined within the angle of  Eq.~\eqref{phimax}. 
{\sf{\textbf{d.}}} Annular phase: the boundaries of the spectral locus are given by the two concentric circles of  Eq.~\eqref{annularradii}.
}
\label{PQCfig3}
\end{figure}

\begin{figure}
\hspace*{1.2 cm}
\includegraphics[width=0.75\linewidth]{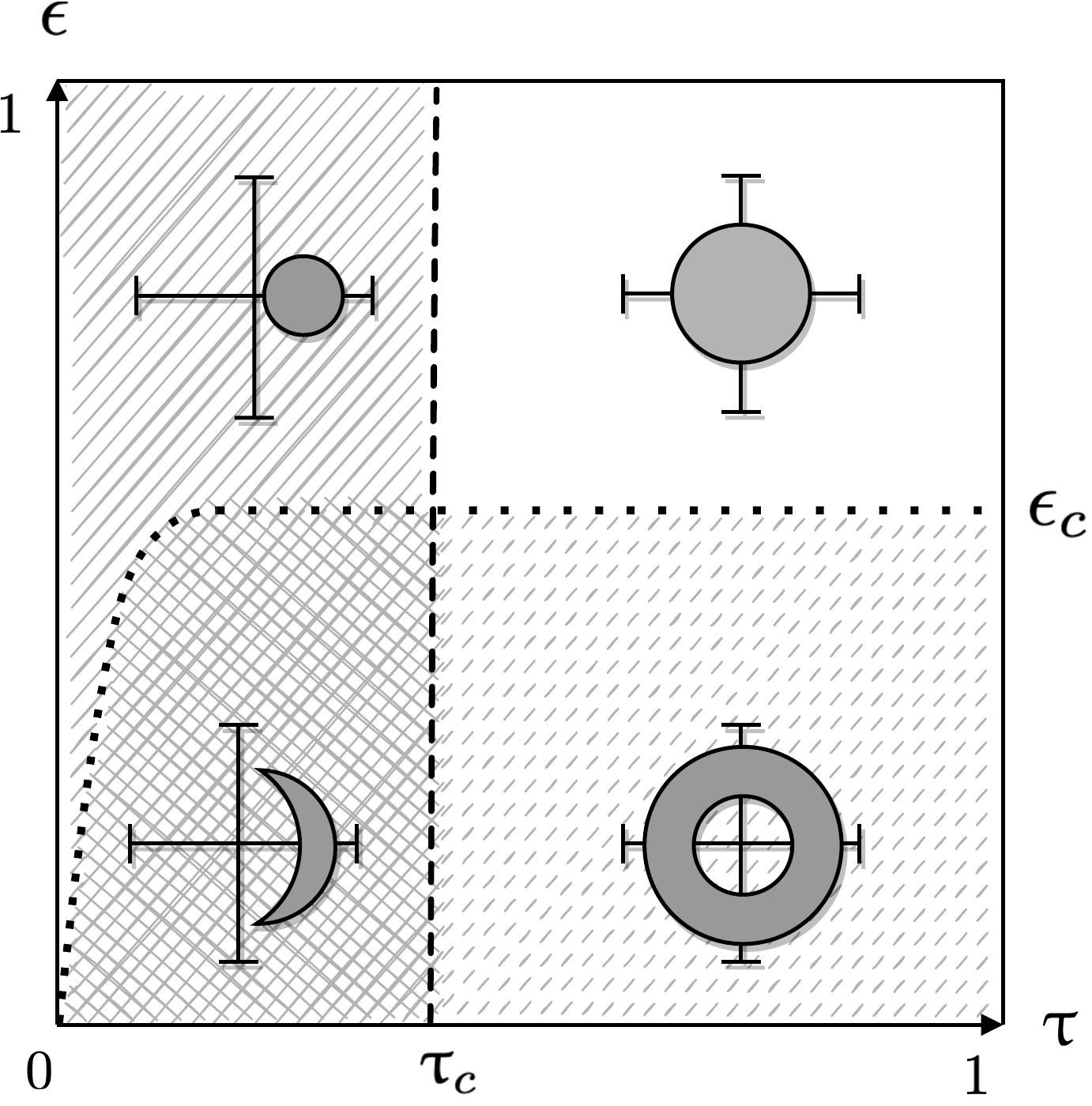}
\caption{Spectral phase diagram of random PQC.
Qualitative classification of the behavior of the spectral bulk for $\hbar = 1$. The dashed lines separate the four main areas of the numerical examples of Fig. \ref{PQCfig3}, as described by  Eq.~\eqref{ecrit}, \eqref{tcrit}, and \eqref{crescentineq}.
}
\label{PQCfig4}
\end{figure}

Let $\lambda \in \mathbb{C}$ be the solutions to the eigenvalue problem
$\mathbb{\Lambda}_{\uptau,\epsilon}   | \lambda ) = \lambda   | \lambda ).$
The eigenvalue $\lambda=1$ is always a solution corresponding to the stationary point of $\mathbb{\Lambda}_{\uptau,\epsilon}$. 
We refer to the rest of the spectrum as the spectral bulk.
The boundaries of the spectral bulk can be categorized into four main classes, as shown in the numerical examples of Fig. \ref{PQCfig3} and the qualitative spectral phase diagram of Fig. \ref{PQCfig4}. 
Specifically, for large values of the dissipation period $\uptau > \uptau_c$ there are two possible spectral phases referred to as "annular" and "disk", a fact known in the literature as the single ring theorem \cite{ginibre1965,girko1985,feinberg2001,molinari2009,tao2010,gotze2010,guionnet2011,fischmann2012,kukulski2021}.
 When a dissipation strength $\epsilon$ is introduced, a phase crossover takes place \cite{sa2020}.  
The decrease of the period $\uptau$ breaks the rotational invariance of the superoperator spectrum, eventually suppressing its bulk within a "shifted disk".  
The crossover from an annulus to a shifted disk happens through an extended "crescent" transitive area.

In the annular phase of Fig. \ref{PQCfig3}{\sf{\textbf{d}}}, the spectrum is bounded by the two concentric circles
\begin{align}
\label{annularradii}
C_{A,\pm} = \sqrt{ (1-\epsilon)^2 \pm \frac{\epsilon^2}{K}}   e^{i \phi}, \quad \phi \in [0,2 \pi],
\end{align}
whose radii can be estimated through a quaternionic resolvent calculation \cite{sa2020}.
The boundary of the disk phase, shown in Fig. \ref{PQCfig3}{\sf{\textbf{b}}}, follows the equation of the outer radius
\begin{align}
\label{dboarder}
C_D = \sqrt{ (1-\epsilon)^2 + \frac{\epsilon^2}{K}}   e^{i \phi} , \quad \phi \in [0,2 \pi]   .
\end{align}
For large dissipation periods, the annulus-to-disk crossover appears at the critical dissipation strength 
\begin{align}
\label{ecrit}
\epsilon_c = \frac{1}{1 + \frac{1}{\sqrt{K}}} ,
\end{align}
i.e., when the argument of the square root of the inner concentric circle in Eq.~\eqref{annularradii} becomes negative \cite{sa2020}, turning the corresponding radius into a purely imaginary number. 
We refer the reader to appendix \ref{apx1} for more details.

As shown in the example of Fig. \ref{PQCfig3}{\sf{\textbf{a}}}, small dissipation periods confine the spectral bulk within the shifted-disk 
\begin{align}
\label{sdboarder}
C_{SD}= 1-\epsilon + \frac{\epsilon}{\sqrt{K}}   e^{i \phi} , \quad \phi \in [0,2 \pi]   ,
\end{align}
result of the angular suppression mechanism explained in more detail in the appendix \ref{apx2}. 
In particular, the argument of the unitary exponential of Eq.~\eqref{model} limits the angular distribution of the spectrum on the complex plane by a $\phi_\mathrm{max} = \uptau   \max\{ E_m - E_n \} /\hbar$.  
Here, we sample the Hamiltonian matrix from $\mathrm{GOE}(d)$, whose spectrum is given by the semicircle law of Eq.~\eqref{semicircle}, resulting in a maximum angle
\begin{align}
\label{phimax}
\phi_\mathrm{max} = \frac{ \uptau}{\hbar}   \sigma \sqrt{8 d}   .
\end{align}

The crossover from an annulus or a disk to a shifted disk roughly starts when all eigenvalues are confined in angles less than $2 \pi$, determining the critical dissipation period $\uptau_c =  \pi \hbar /(\sigma \sqrt{2 d}) $, which is related to the Heisenberg time of the ED models \eqref{RMTheisenberg} through
\begin{align}
\label{tcrit}
 t_\text{H} = (d-1) \uptau_c .
\end{align}
This result is not restricted to RMT but applies whenever disordered averages are considered, given that the dimension of the Hilbert space is finite and the spectrum of the Hamiltonian is bounded. 

For $\epsilon < \epsilon_c$, the crossover to a shifted disk goes through the transitive crescent area of Fig. \ref{PQCfig3}{\sf{\textbf{c}}}, ending when the imaginary part of all bulk eigenvalues becomes less than the radius of the shifted disk
\begin{align}
\label{crescentineq}
\epsilon \leq \frac{1}{1 + \frac{1}{\sqrt{K} \sin ( \frac{ \uptau}{\hbar}   \sigma \sqrt{8 d} )}} .
\end{align}
We note that Eq.~\eqref{crescentineq} is valid until the critical dissipation period $\uptau_c$, after which it reduces to Eq.~\eqref{ecrit}.

We note that knowledge of the spectral properties of the channel of Eq. \eqref{supermodel} is sufficient to construct those of the time-evolved channel in Eq. \eqref{supermodelpower}, as explained in Appendix \ref{apx4}.

\subsection{Correlation Hole}

\begin{figure}
\hspace*{0 cm}
\includegraphics[width=0.95\linewidth]{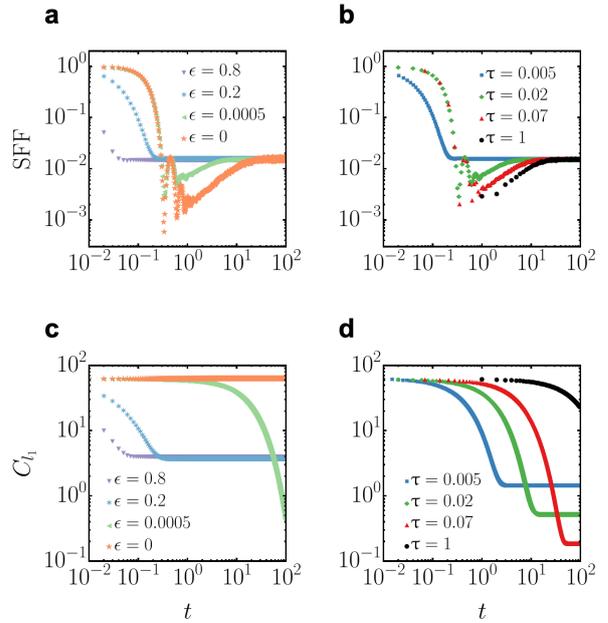}
\caption{Shrinking of the correlation hole.
The infinite temperature SFF, $\beta=0$ at time $t = j \uptau$, evolved through the consecutive applications of a random PQC \eqref{supermodelpower}. 
In panels {\sf{\textbf{a.}}} and {\sf{\textbf{c.}}} we show the shrinking of the correlation hole and the corresponding decay of the CGS $l_1$-norm of coherence, respectively, for constant dissipation period $\uptau = 0.01$ and increasing $\epsilon$. 
In panels {\sf{\textbf{b.}}} and {\sf{\textbf{d.}}} we show the shrinking of the correlation hole and the corresponding decay of the CGS $l_1$-norm of coherence, respectively, for constant dissipation strength $\epsilon = 0.01$ and decreasing $\uptau$.
We have taken the average fidelity in all panels over a sample of $500$ $\mathrm{GOE}(64)$, $\sigma =1$ random Hamiltonian matrices, setting $\hbar = 1$.
The corresponding sets of $K=3$ random Kraus-Choi matrices $N_r$ were drawn as truncations of an enlarged random $\mathrm{CUE}(192)$ element. 
}
\label{PQCfig5}
\end{figure}

The existence of a correlation hole in the structure of an auto-correlation function, such as the SFF of Eq.~\eqref{sffopen}, is a signature of the underlying complexity, reflecting the level of commensurability between the frequencies associated with the dynamics \cite{leviandier1986,wilkie1991,alhassid1992,ma1995,gorin2006,torres_herrera2017}.
In the energy eigenbasis of a subsystem, the contribution of these frequencies to the time evolution is weighted by the size of the corresponding coherences of the density matrix, i.e., by the size of its off-diagonal elements.
Said differently, starting from an initially coherent state (as defined by the resource theory of coherence \cite{streltsov2017}), the relative phases of the probability amplitudes of the energy eigenstates vanish as the state becomes incoherent.
The suppression of coherences due to the dynamical map's non-unitarity is thus expected to suppress the correlation hole's size.

In Sec. \ref{SecED}, we illustrated the mechanism underlying the dynamical suppression of the correlation hole in the paradigmatic case of ED, interpreting the results of \cite{xu2021,Cornelius22,matsoukas_roubeas2023} through the resource theory of coherence \cite{streltsov2017}.
There, the maximally incoherent operation of the ED channel turns the maximally coherent CGS at infinite temperature asymptotically into an incoherent one.
The random PQC is incoherent but not necessarily maximally incoherent. 
Thus, the $l_1$-norm of coherence does not saturate asymptotically to zero.
This results in loosening the bounds of Eq.~\eqref{sffleqcl1} when we increase the dissipation strength or decrease the dissipation period, while the saturation time of the SFF and the $l_1$-norm of coherence is shifted to the left of $t_\text{H}$ before the correlation hole closes.
In Fig. \ref{PQCfig5}, we show the corresponding shrinking of the correlation hole with the increment of the dissipation strength or the decrement of the dissipation period for random PQC.  

\begin{figure}
\hspace*{0.8 cm}
\includegraphics[width=0.73\linewidth]{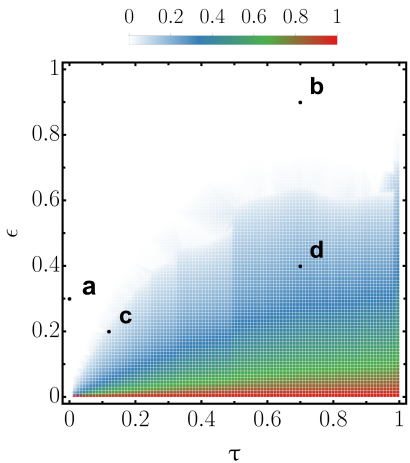}
\caption{ Relative effective depth of the correlation hole. 
The infinite temperature $\text{SFF}(j \uptau)$ for the channel $\mathbb{\Lambda}_{\uptau,\epsilon}^j$ is averaged over $100$ Hamiltonians $H$ sampled from $\mathrm{GOE}(64)$, with $\sigma =1$, setting $\hbar = 1$. 
The corresponding $100$ sets of $K=3$ random Kraus-Choi matrices $N_r$ are drawn as truncations of an enlarged random $\mathrm{CUE}(192)$ element. 
The points {\sf{\textbf{a}}} , {\sf{\textbf{b}}} , {\sf{\textbf{c}}}  and {\sf{\textbf{d}}} correspond to the four panels in Fig. \ref{PQCfig3},\ref{PQCfig7}, and \ref{PQCfigapx3b}. 
}
\label{PQCfig6}
\end{figure}

We define the effective depth of the correlation hole
\begin{align}
\label{effdepthPQC}
D_\text{eff} = \sqrt{ \ln \displaystyle \prod_{j=j_\text{Th}}^{j_\text{H}} \frac{F_p}{\text{SFF}(j \uptau)} }  .
\end{align}
The time steps of the Thouless and Heisenberg times are defined through the ceiling functions $j_\text{Th} = \lceil t_\text{Th}/ \uptau \rceil$, $j_\text{H} = \lceil t_\text{H}/ \uptau \rceil$.
The relative effective depth can be defined as the effective depth normalized by the corresponding value for the isolated dynamics.
In Fig. \ref{PQCfig6}, we show the relative effective depth of the correlation hole in the plane of the PQC parameters.
See also Fig. \ref{PQCfigapx3b} in appendix \ref{apx3} for an illustration of the SFF with specific choices of the PQC parameters.

\subsection{Complex Spacing Ratios}

\begin{figure}
\hspace*{0.2 cm}
\includegraphics[width=1\linewidth]{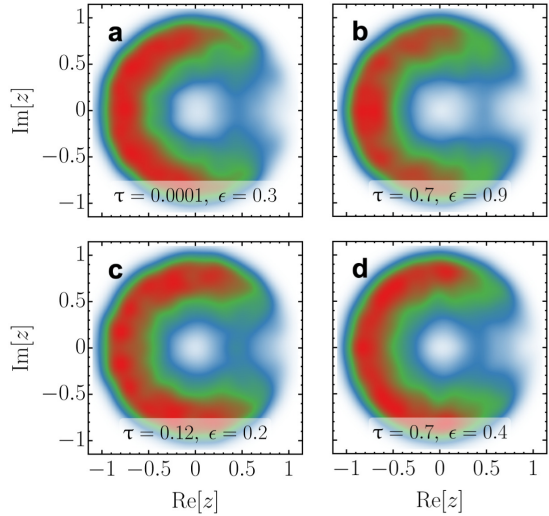}
\caption{Complex Spacing Ratios of random PQC.
Density plot on the complex plane (blue to red) of $16384$ complex spacing ratios corresponding to the spectral densities of panels {\sf{\textbf{a}}}, {\sf{\textbf{b}}}, {\sf{\textbf{c}}}  and {\sf{\textbf{d}}} of Fig.  \ref{PQCfig3} and \ref{PQCfigapx3b} respectively.  
Each panel was generated by $4$ independent realizations,  with the Hamiltonian $H$ sampled from $\mathrm{GOE}(64)$, $\sigma =1$, setting $\hbar = 1$ and a set of $K=3$ random Kraus-Choi matrices $N_r$, drawn as truncations of an enlarged random $\mathrm{CUE}(192)$ element.  
The color scheme and scaling are the same as that of Fig. \ref{PQCfig3} and \ref{PQCfig6}.
}
\label{PQCfig7}
\end{figure}

In isolated dynamics, the relation between the static signatures of quantum chaos and the corresponding dynamical ones has been explored.
In particular, the rigidity of the Hamiltonian spectrum has been related to the existence of a linear ramp in the correlation hole after the Thouless time $t_\text{Th}$ and before its saturation to the plateau at $t_\text{H}$.
However, the spectrum of the generators of open quantum dynamics or the corresponding dynamical maps is, in principle, spread around the complex plane.
Hence, the generalization of signatures involving level spacing distributions is not straightforward.
One possibility introduced in \cite{sa2020b} is the following.
Given an eigenvalue $\lambda$ of $\mathbb{\Lambda}_{\uptau,\epsilon}$ on the complex plane,  one can find the nearest,  $\lambda^{NN}$ and next nearest,  $\lambda^{NNN}$ neighboring eigenvalues respectively,  and thus define the complex spacing ratio $z = ( \lambda^{NN} - \lambda)/(\lambda^{NNN} - \lambda )$.

As we saw in the previous sections, the shrinking of the correlation hole of the SFF is a purely dynamical effect linked to the behavior of the corresponding coherence monotones of the CGS. 
Static signatures of open quantum chaos, such as complex spacing ratios, do not necessarily diagnose the long-time behavior of the coherences in a direct way.
In Fig. \ref{PQCfig7}, we show the density plots of the complex spacing ratios for PQC parameters corresponding to the four spectral phases of Fig.   \ref{PQCfig3}.
We see that the angular and the radial repulsion persist in all parameter areas, even where the correlation whole has vanished.

\section{Summary}

Known signatures of quantum chaos, coherence, and Markovianity are measures of quantum correlations.
Based on this observation, our work proposes an information-theoretic framework to explore the interplay between environmental noise and complexity in quantum systems. 
By relating the survival probability of an initial CGS to the associated $l_1$-norm of coherence, we account for complex dissipative and not-necessarily-Markovian quantum effects.
The manifestations of correlations between energy eigenvalues associated with quantum chaos, which result in the correlation hole of the SFF, are suppressed by non-unitary dynamics during the time evolution. 
This suppression is a direct consequence of the loss of coherence in the energy eigenbasis, measured by the different coherence monotones.
Nevertheless, such dynamical phenomena are hard to capture by the proposed static spectral signatures of dissipative quantum chaos.
We work on paradigmatic examples of maximally incoherent Lindbladian dynamics and a discrete-time model of CPTP maps involving a unitary generated by a Hamiltonian. 
The latter is interesting in its own right. We have thus discussed its spectral properties, sampling the constituting operations by random matrix ensembles.

Our work provides a natural and tractable framework to explore the interplay between 
environmental noise, resource theories and complexity in scalable quantum systems. 
This quest is of particular importance for the transition from the NISQ era in quantum technologies \cite{preskill2018}.
Finally, our work contributes to understanding the emergence of classical behavior from the quantum substrate, e.g., the quantum-to-classical transition \cite{zurek2003}. In particular, we have shown that signatures of Hamiltonian quantum chaos rely on the presence of quantum coherence and are thus suppressed by dephasing mechanisms stemming from decoherence. These findings are complementary to the pioneering works exploring the interplay of quantum chaos and decoherence that relied on the latter for the emergence of classical chaos \cite{ZurekPaz94,Karkuszewski02,zurek2003,habib2005}. 

\begin{acknowledgments}
It is a pleasure to acknowledge useful discussions with Pablo Martínez-Azcona and Simon Lieu. 
This project was supported by the Luxembourg National Research Fund (FNR Grant No. 16434093). It has received funding from the QuantERA II Joint Programme with co-funding from the European Union’s Horizon 2020 research and innovation programme. 
\end{acknowledgments}

\bibliography{PQCquantum}
\bibliographystyle{quantum}


\bigskip \smallskip
\hrule \enskip \hrule

\section*{Appendix}
\appendix 
\section{Annulus-to-Disk Crossover}
\label{apx1}

\begin{figure}[H]
\hspace*{-0.5 cm}
\includegraphics[width=1 \linewidth]{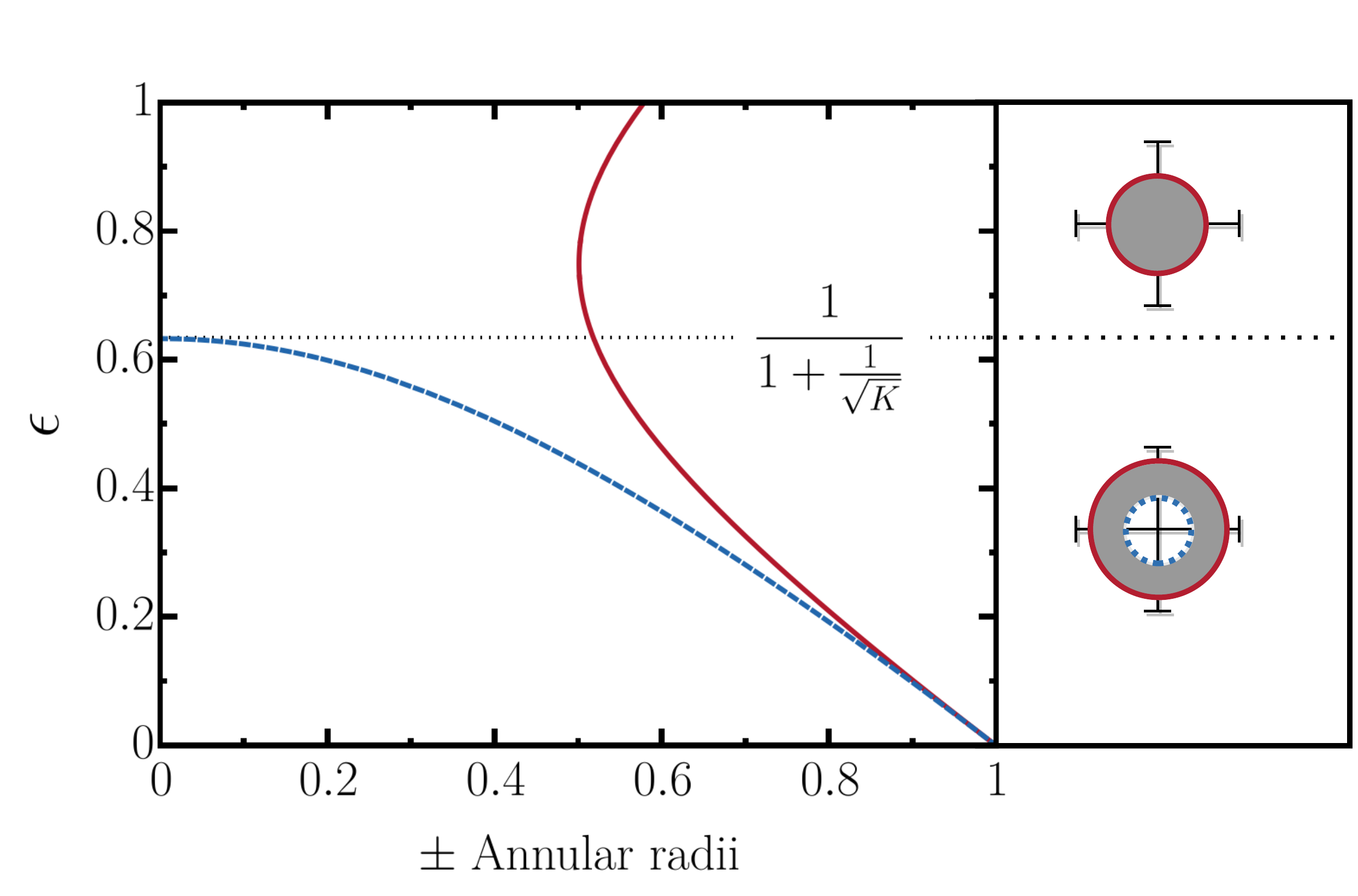}
\caption{Annulus-to-disk crossover.
The crossover of the spectrum from an annulus to a disk is manifested in the collapse of the inner radius (dashed blue line) of Eq.~\eqref{annularradii} into a complex number.  
After the crossover, the outer radius (solid red line) becomes the disk's radius.
For illustrative purposes, we plot the example of $K=3$, corresponding to a critical dissipation strength $\epsilon_c \simeq 0.634$.
}
\label{PQCfigapx1}
\end{figure}

The circular law is a fundamental result of RMT, stating that the distribution of eigenvalues of square random matrices, with independent and identically distributed complex entries, in the limit of infinite dimension is uniform over a disk \cite{ginibre1965, girko1985, gotze2010, tao2010}.
The single ring theorem concerns the confinement of the eigenvalue distribution of a large class of non-Hermitian random matrices within a disk or an annulus \cite{feinberg2001,guionnet2011,bruzda2009,bruzda2010,fischmann2012,kukulski2021}.
The annulus-to-disk spectral crossover of the PQC follows directly from the study of the corresponding $1$-parameter CPTP maps \cite{sa2020}.
In this case, the critical dissipation strength can be calculated for an effective model using non-Hermitian free probability.
 Fig. \ref{PQCfigapx1} shows an example of the inner radius collapse. 

\section{Angular Suppression and Shift of the Superoperator Spectrum}
\label{apx2}

The angular suppression of the superoperator spectrum of Eq.~\eqref{supermodelpower} originates in the finite range of the Hamiltonian eigenvalue distribution.  
In this section, we discuss how the maximum energy gap can confine all superoperator eigenvalues in a circular sector of the complex plane, resulting in the spectral phase crossover to a shifted disk.

\begin{figure}
\hspace*{-0.2 cm}
\includegraphics[width=1 \linewidth]{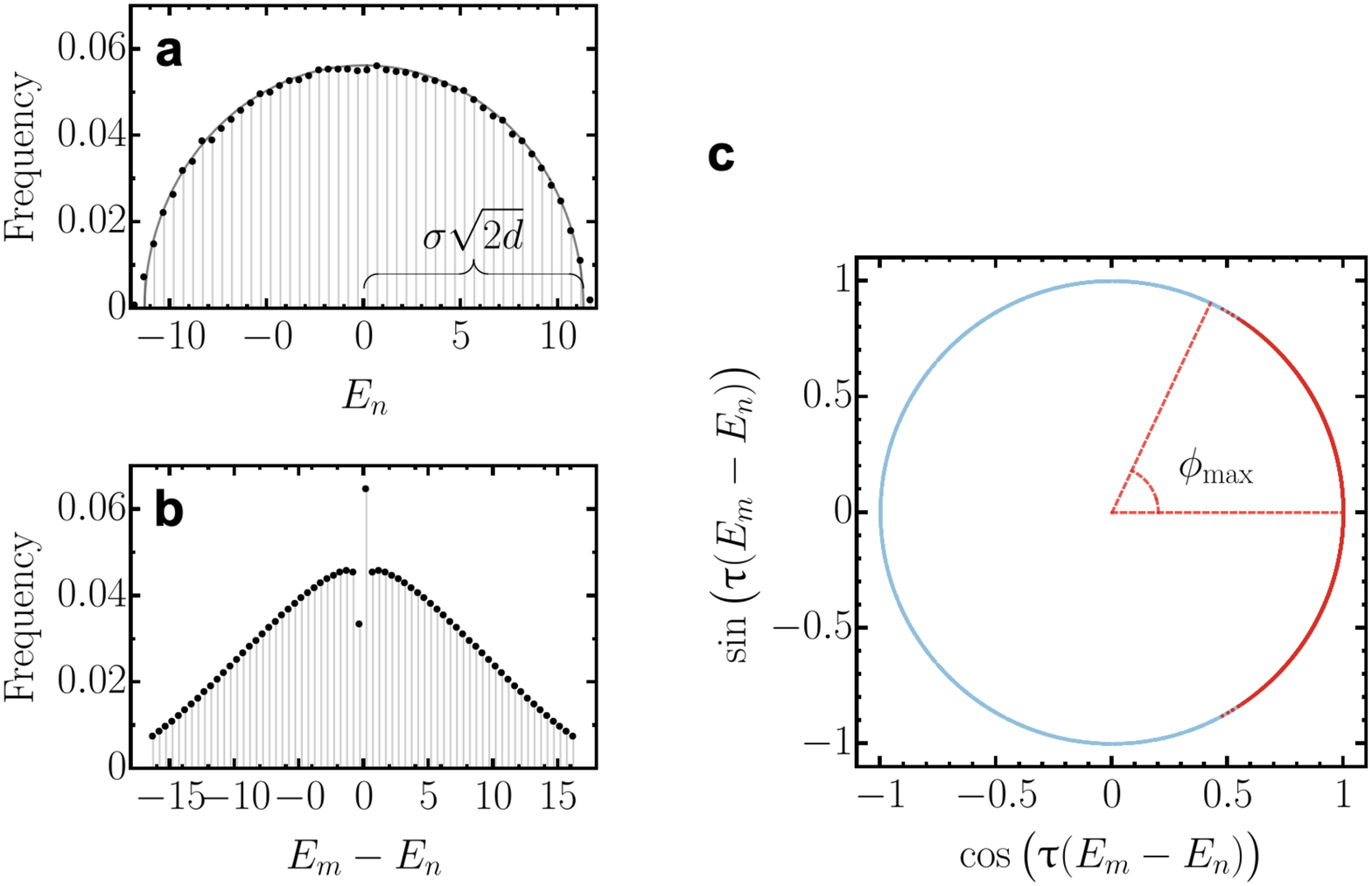}
\caption{Angular suppression of the superoperator spectrum. 
{\sf{\textbf{a.}}} Histogram of the eigenvalues of $10^4$ Hamiltonians and the corresponding semicircle law. 
{\sf{\textbf{b.}}} Histogram of the eigenvalues of $10^4$ Liouvillians, i.e. all energy gaps. 
{\sf{\textbf{c.}}} Spectrum of a $\exp{-i \frac{\uptau}{\hbar}    H} \otimes   \exp{ i \frac{\uptau}{\hbar}   H}$ for $\uptau = 1$ (blue) $\uptau = 0.05$ (red).
The Hamiltonians $H$ were sampled from $\mathrm{GOE}(64)$, $\sigma =1$, setting $\hbar = 1$.
}
\label{PQCfigapx2a}
\end{figure}

Let us first illustrate how the suppression mechanism breaks the rotational invariance of the spectrum when $\epsilon = 0$, i.e., when our system is isolated from the environment. 
In this case, the Hamiltonian operator is Hermitian, the superoperator $ \mathbb{\Lambda}_{\uptau,\epsilon}^j$ is unitary, and its eigenvalue problem becomes
\begin{align}
\mathbb{\Lambda}_{\uptau,\epsilon}^j   ( \ket{n} \otimes \ket{m} ) 
&= e^{i \frac{j \uptau}{\hbar}   (\id \otimes H^\intercal - H \otimes \id)}   ( \ket{n} \otimes \ket{m} ) \nonumber \\
&= e^{i \frac{j \uptau}{\hbar}   (E_m - E_n)}   ( \ket{n} \otimes \ket{m} )   .
\end{align}
Then, the spectrum belongs in a circular sector of central angle $2 \phi_\mathrm{max}$, where $\phi_\mathrm{max} = j \uptau   \max\{ E_m - E_n \} / \hbar$. 
 In Fig. \ref{PQCfigapx2a}{\sf{\textbf{c}}}, we show the example of a Hamiltonian sampled from $\mathrm{GOE}(64)$ for two different values of $\uptau$ and $j=1$.
There, the largest energy difference is given by the semicircle \eqref{semicircle} radius, $\max\{ E_m - E_n \} = \sigma \sqrt{8 d}$, resulting in a maximum angle, $\phi_\mathrm{max} = j \uptau   \sigma \sqrt{8 d} / \hbar$.

\begin{figure}
\hspace*{-0.2 cm}
\includegraphics[width=1\linewidth]{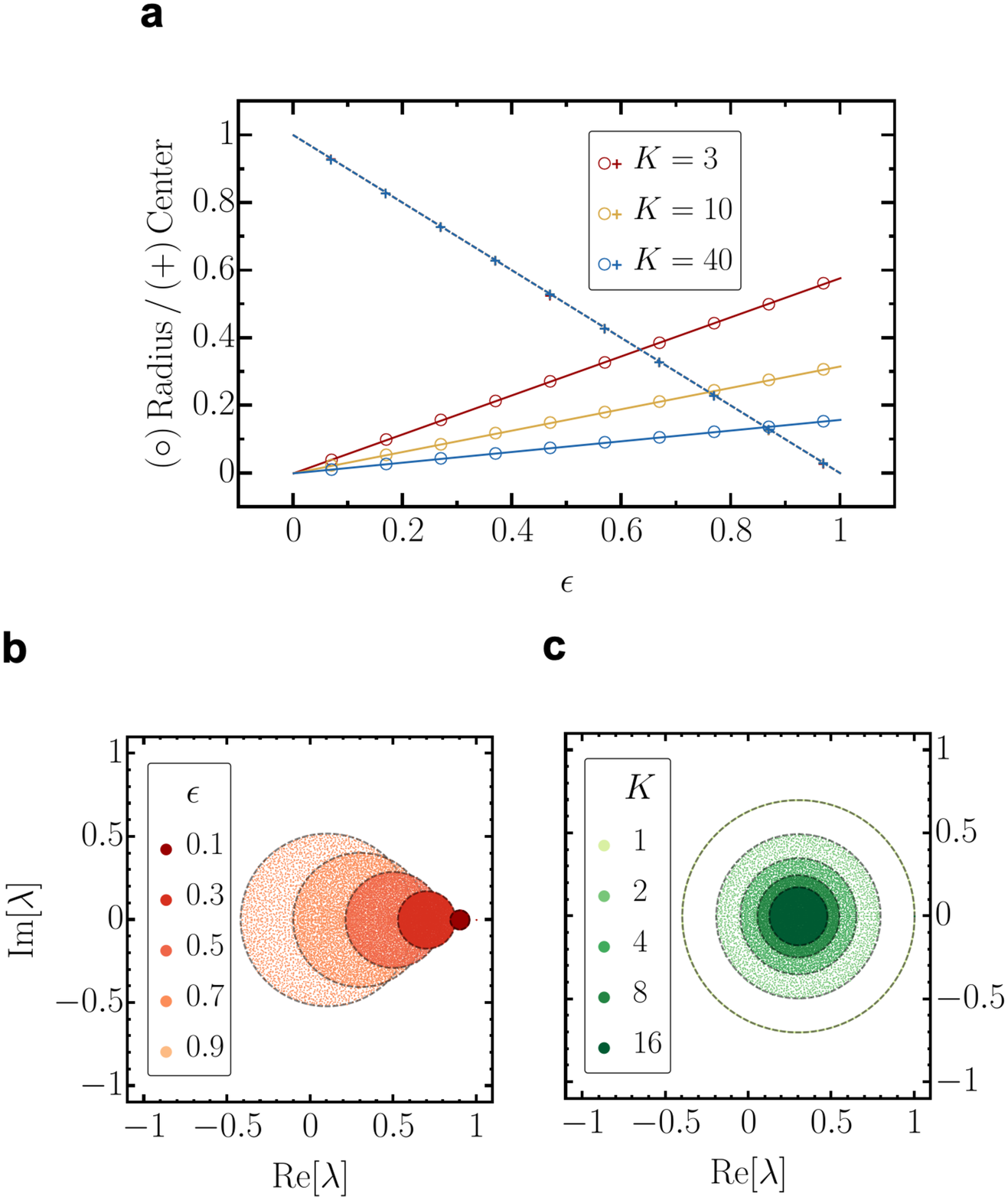}
\caption{Shifted-disk phase. 
{\sf{\textbf{a.}}} Radii ($\circ$, solid colored lines) and centers ($+$, dashed colored lines) of the border of the spectral loci, together with the corresponding analytic prediction of Eq.~\eqref{sdboarder}, for different numbers of Kraus-Choi operators $K$.
{\sf{\textbf{b.}}} Spectra on the complex plane for $K=3$ and varying $\epsilon$.
{\sf{\textbf{c.}}} Spectra on the complex plane for $\epsilon=0.7$ and varying $K$.  
The dashed black circles are the theoretical border of Eq.~\eqref{sdboarder}.
For every set of parameters, we show the eigenvalues of a single Hamiltonian $H$ sampled from $\mathrm{GOE}(64)$, $\sigma =1$, setting $\hbar = 1$ and $\uptau=0.0001$, while a set of $K$ random Kraus-Choi matrices $N_r$ are drawn as truncations of an enlarged random $\mathrm{CUE}(192)$ element. } 
\label{PQCfigapx2b}
\end{figure}

When the periodic interaction with an environment is introduced by increasing $\epsilon$, the above mechanism gradually shifts the real part of the overall superoperator spectrum to the right of the complex plane.  
For time-independent systems, the Hamiltonian eigenvalues are fixed, and the only parameter controlling the spectral crossover to a shifted-disk phase is the dissipation period $\uptau$.  
For decreasing $\uptau$, the crossover to a shifted disk starts roughly when $\phi_\mathrm{max} \leq 2 \pi$, and ends when the characteristic radius of the spectral locus of the dissipative term in the quantum channel becomes of the order of $\sin(\phi_\mathrm{max})$, as indicated by Eq.~\eqref{crescentineq}. 
After this point, the spectrum asymptotically approaches the one of the dissipative part, shifted to the right by $1-\epsilon$, since the overall channel for $\uptau \rightarrow 0$ is 
\begin{align}
\thinmuskip=0mu
\mathbb{\Lambda}_{ \uptau \rightarrow 0,\epsilon}  \rightarrow (1-\epsilon)   \id \otimes \id + \epsilon   \sum_{r=1}^{K} N_r^{\phantom *} \otimes N_r^*   .
\end{align}
If the dissipative part is composed of a set of random Kraus-Choi operators, as in the RMT examples of section \ref{SecRPQC}, the shifted spectrum belongs in a disk of radius given by the inverse of the square root of the number of the Kraus-Choi operators $\epsilon/\sqrt{K}$  \cite{sa2020}, a result that can also easily be obtained by taking the limit $\epsilon \rightarrow 1$ in Eq.~\eqref{dboarder}.  
In Fig. \ref{PQCfigapx2b}, we show the agreement of the numerical calculations with the prediction of Eq.~\eqref{sdboarder}.

\section{Shrinking of the Correlation Hole}
\label{apx3}

\begin{figure}
\hspace*{-0.3 cm}
\includegraphics[width=1.06 \linewidth]{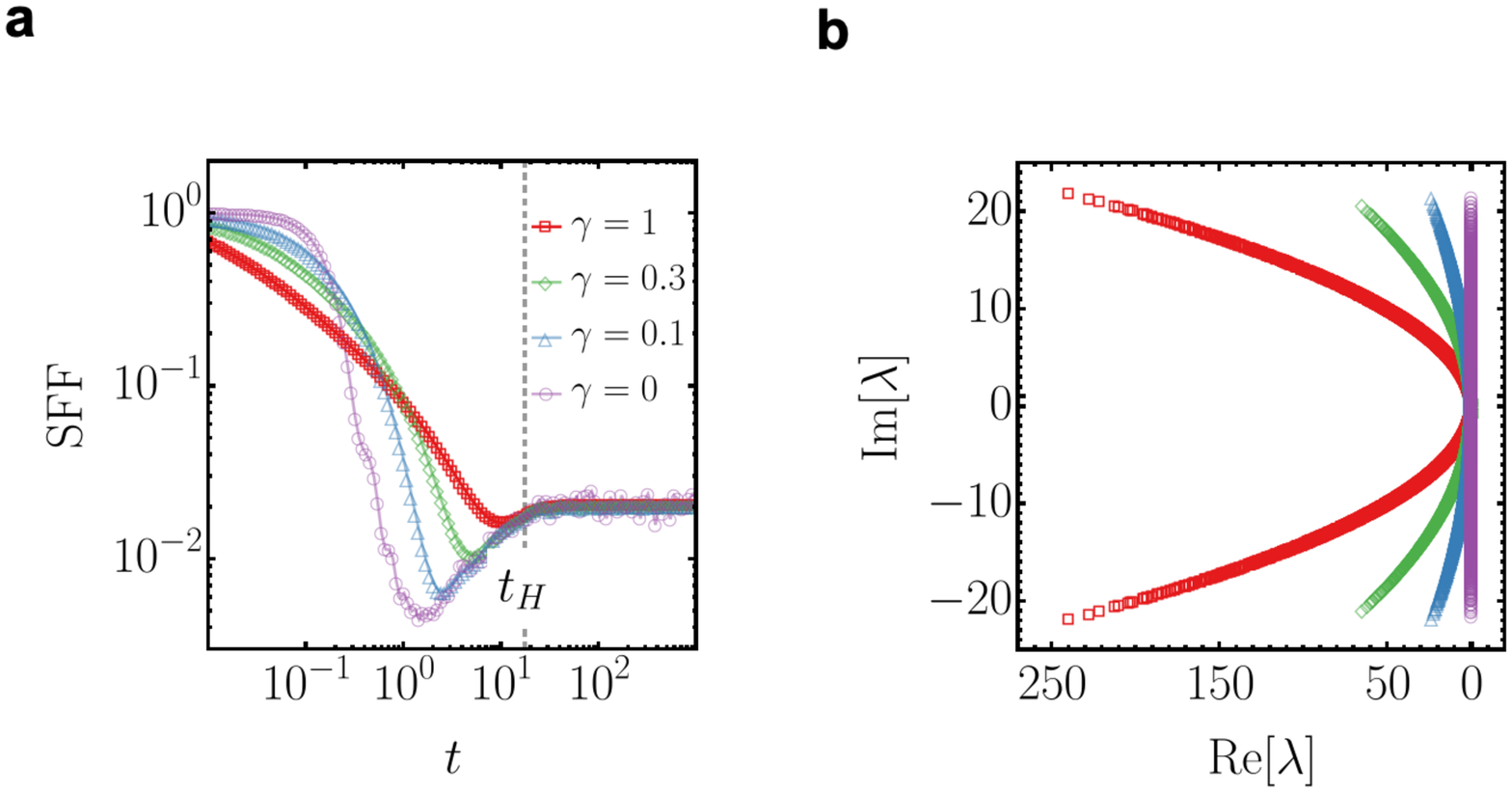}
\caption{Shrinking of the correlation hole in ED dynamics. 
{\sf{\textbf{a.}}} Hamiltonian average of the SFF for different dephasing strengths. 
We consider a sample of $100$ $\mathrm{GOE}(64)$, $\sigma=1$ matrices, at inverse temperature $\beta=0.1$. 
We have set $\hbar=1$.
{\sf{\textbf{b.}}} Spectra of the corresponding Liouvillians generating the dynamics.
}
\label{PQCfigapx3a}
\end{figure}

As we have seen, the suppression of the correlation hole in the survival probability of an initial CGS is related to the suppression of coherence in the evolved state, a phenomenon that cannot be straightforwardly diagnosed in the spectral structure of the dynamical generators.
Fig. \eqref{PQCfigapx3a} shows the correlation hole of the averaged SFF closing next to the corresponding Liouvillian spectra by increasing the ED dissipation strength of Eq.~\eqref{dyned}.
This is a case of maximally incoherent operation leading to the complete vanishing of the density matrix's off-diagonal elements in the energy eigenbasis.
The Heisenberg time remains unchanged, while the Thouless time is shifted to the right.

\begin{figure}
\hspace*{-0.2 cm}
\includegraphics[width=0.9 \linewidth]{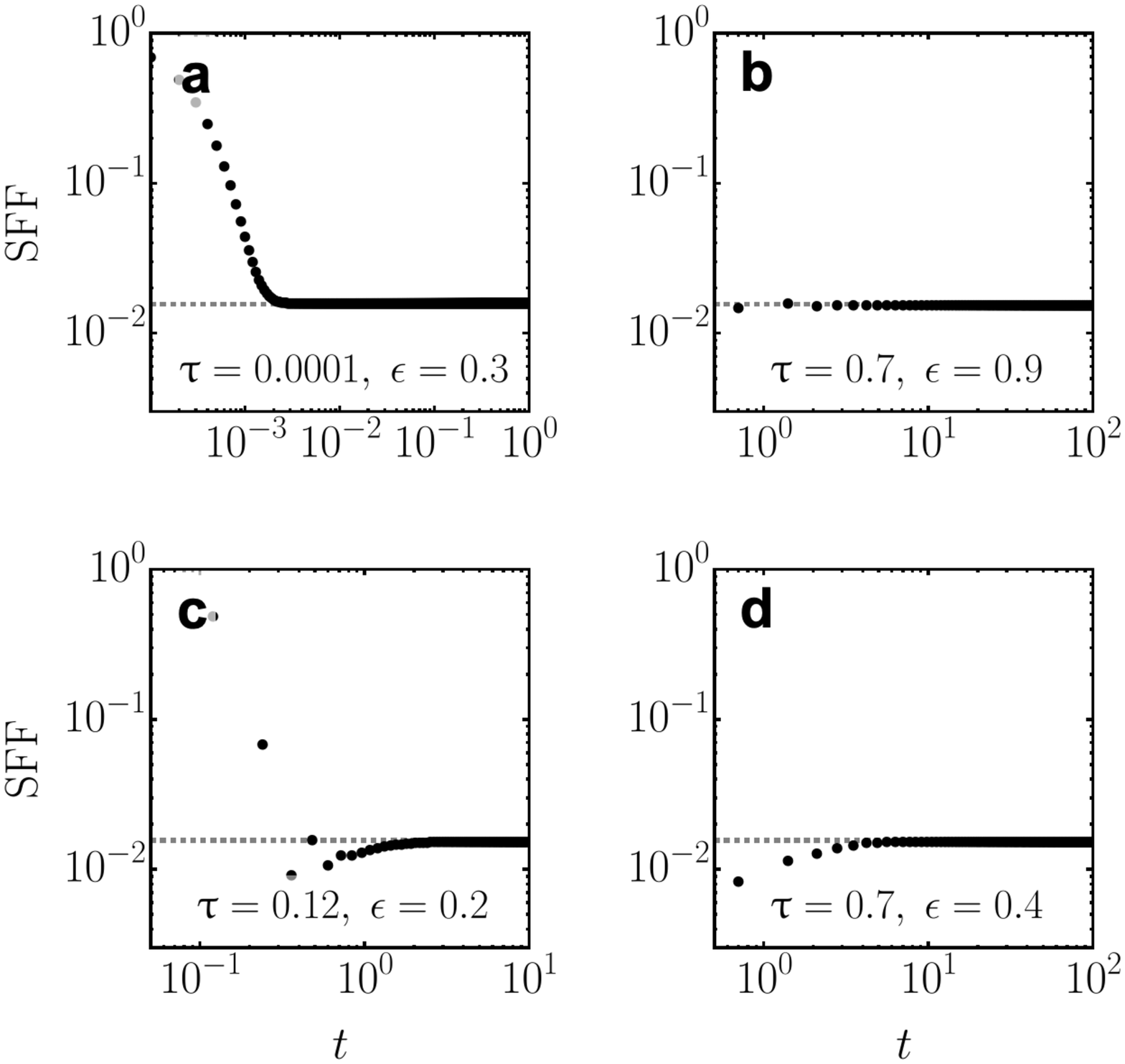}
\caption{Averaged SFF corresponding to the spectral densities of panels {\sf{\textbf{a}}} , {\sf{\textbf{b}}} , {\sf{\textbf{c}}}  and {\sf{\textbf{d}}} of Fig.  \ref{PQCfig3},\ref{PQCfig7} respectively.  
The relative effective depth of the correlation holes for the specific values of the parameters $\uptau$ and $\epsilon$ is shown in Fig. \ref{PQCfig6}.
In all panels, we have taken the average fidelity over a sample of $100$ $\mathrm{GOE}(64)$, $\sigma =1$ random Hamiltonian matrices, setting $\hbar = 1$.
The corresponding sets of $K=3$ random Kraus-Choi matrices $N_r$ were drawn as truncations of an enlarged random $\mathrm{CUE}(192)$ element. 
}
\label{PQCfigapx3b}
\end{figure}

Fig. \ref{PQCfigapx3b} shows the correlation hole of the averaged SFF closing by increasing the dissipation strength $\epsilon$ or decreasing the dissipation period $\uptau$ for the discrete PQC evolution in Eq.~\eqref{supermodelpower}.
In this case, the Heisenberg time is shifted to the left while the Thouless time remains unchanged.

\section{Time Evolution of the PQC Spectrum}
\label{apx4}

\begin{figure}
\hspace*{0 cm}
\includegraphics[width=1\linewidth]{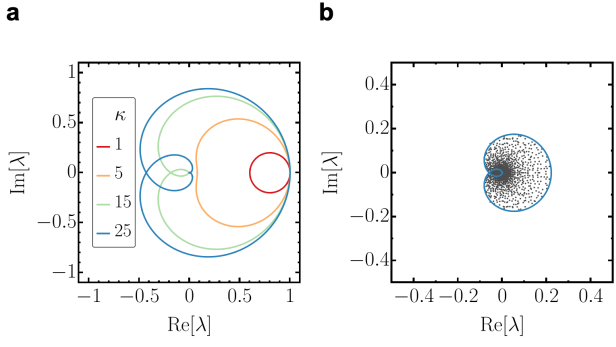}
\caption{Time evolution of the PQC spectrum.
{\sf{\textbf{a.}}} Different powers $\kappa$ of the spectral boarder of the channel $\mathbb{\Lambda}_{\uptau,\epsilon}$ in the shifted-disk phase.
{\sf{\textbf{b.}}} Eigenvalues (grey points) and border (blue line) of the superoperator $\mathbb{\Lambda}_{\uptau,\epsilon}^\kappa$ with $\kappa = 25$ and $\epsilon=0.2$.  
Single realization with the Hamiltonian $H$ sampled from $\mathrm{GOE}(64)$, $\sigma =1$, setting $\hbar = 1$ and $\uptau=0.0001$,  the set of $K = 2$ random Kraus-Choi matrices $N_r$ are drawn as two sequential blocks of a random $\mathrm{CUE}(192)$ element.  
}
\label{PQCfigapx4}
\end{figure}

The powers of the eigenvalues of a matrix are the eigenvalues of the corresponding matrix powers.
Accordingly, the spectral loci of the different powers of the PQC \eqref{supermodelpower} are bounded by the powers of Eq.~\eqref{annularradii}, \eqref{dboarder} and \eqref{sdboarder}, for the annular, the disk and the shifted disk respectively.
In the case of the annular and disk phases, they result in concentric circles, while in the shifted-disk case, they form the cardioids of Fig. \ref{PQCfigapx4}.

\section{Proof of the $C_{l_1}$ bound to the SFF}
\label{apx5}

The fidelity between an initial CGS and its time evolution can be written as
\begin{align} 
{\rm SFF}(t) &= \bra{\Psi_\beta}   \rho_\beta (t)  \ket{\Psi_\beta} 
= \tr \big[ \rho_\beta \rho_\beta (t) \big] \\
&= \sum_{n,m=1}^d \sqrt{p_n p_m}  \tr \big[ \ketbra{n}{m}  \rho_\beta (t)   \big] \\ 
&= \sum_{n=1}^d  p_n \rho_{\beta, nn} (t) + \sum_{\substack{n,m=1 \\ n \neq m}}^d  \sqrt{p_n p_m} \rho_{\beta, nm} (t), 
\end{align}

which at infinite temperature $\beta=0$ becomes
\begin{align} 
{\rm SFF}(t) &=  \sum_{n=1}^d  \frac{1}{d} \rho_{0, nn} (t) + \sum_{\substack{n,m=1 \\ n \neq m}}^d  \frac{1}{d} \rho_{0, nm} (t) \\
&= \frac{1}{d} \left( 1 +  \sum_{\substack{n,m=1 \\ n \neq m}}^d \rho_{0, nm} (t) \right) ,
\end{align}
and thus we arrive to Eq.~\eqref{sffopenbeta0}
\begin{align} 
{\rm SFF}(t) = \frac{1}{d} \left( 1 +  2 \sum_{\substack{n,m=1 \\ m < n}}^d \Re[\rho_{0, nm} (t)] \right) .
\end{align}

We now remind Eq.~\eqref{cl1} 
\begin{equation}
C_{l_1} (t) =  2 \sum_{\substack{\ell,k=1 \\ k < \ell}}^d \abs{ \rho_{0,\ell k} (t) }    ,
\end{equation}
and notice that $ -\abs{\rho_{0, nm} (t)} \leq \Re[\rho_{0, nm} (t)] \leq \abs{\rho_{0, nm} (t)} $, which by the addition property of inequalities gives
\onecolumn
\begin{align} 
- \sum_{\substack{n,m=1 \\ n \neq m}}^d \abs{\rho_{0, nm} (t)} 
& \leq  \sum_{\substack{n,m=1 \\ m < n}}^d \Re[\rho_{0, nm} (t)] 
\leq \sum_{\substack{n,m=1 \\ n \neq m}}^d \abs{\rho_{0, nm} (t)}
\Leftrightarrow \\
\frac{1}{d} \left( 1 - 2 \sum_{\substack{n,m=1 \\ n \neq m}}^d \abs{\rho_{0, nm} (t)} \right) 
\leq & \frac{1}{d} \left( 1 +  2 \sum_{\substack{n,m=1 \\ m < n}}^d \Re[\rho_{0, nm} (t)] \right) 
\leq \frac{1}{d} \left( 1 + 2 \sum_{\substack{n,m=1 \\ n \neq m}}^d \abs{\rho_{0, nm} (t)} \right) 
\Leftrightarrow \\
\frac{1}{d} &\big( 1 - C_{l_1} (t) \big)
\leq  {\rm SFF}(t) 
\leq \frac{1}{d} \big( 1 + C_{l_1} (t) \big) .
\end{align}

\end{document}